# Dual-comb cavity ring-down spectroscopy

D. Lisak[1*†], D. Charczun[1†], A. Nishiyama[1], T. Voumard[2], T. Wildi[2], G. Kowzan[1], V. Brasch[3], T. Herr[2], A. J. Fleisher[4], J. T. Hodges[4], R. Ciuryło[1], A. Cygan[1], P. Masłowski[1]

1) Institute of Physics, Faculty of Physics, Astronomy and Informatics, Nicolaus Copernicus University in Toruń, Grudziądzka 5, 87-100 Toruń, Poland
2) Center for Free-Electron Laser Science (CFEL), German Electro-Synchrotron (DESY), Notkestr. 85, 22607 Hamburg, Germany
3) CSEM - Swiss Center for Electronics and Microtechnology, 2000 Neuchâtel, Switzerland
4) Optical Measurements Group, National Institute of Standards and Technology, 100 Bureau Drive, Gaithersburg, Maryland 20899, U.S.A.

* dlisak@umk.pl
† equal contributions
June 14, 2021

**Abstract**

Cavity ring-down spectroscopy is a ubiquitous optical method used to study light-matter interactions with high resolution, sensitivity and accuracy. However, it has never been performed with the multiplexing advantages of direct frequency comb spectroscopy without sacrificing orders of magnitude of resolution. We present dual-comb cavity ring-down spectroscopy (DC-CRDS) based on the parallel heterodyne detection of ring-down signals with a local oscillator comb to yield absorption and dispersion spectra. These spectra are obtained from widths and positions of cavity modes. We present two approaches which leverage the dynamic cavity response to coherently or randomly driven changes in the amplitude or frequency of the probe field. Both techniques yield accurate spectra of methane — an important greenhouse gas and breath biomarker. The high sensitivity and accuracy of broadband DC-CRDS, shows promise for applications like studies of the structure and dynamics of large molecules, multispecies trace gas detection and isotopic composition.

**Introduction**

With the availability of tunable, narrow-linewidth lasers and high-reflectivity dielectric mirrors, cavity ring-down spectroscopy (CRDS) [1] is one of the most widely used, accurate and sensitive spectroscopic techniques. Example applications include studies of fundamental interactions, atmospheric composition, dynamics, and radiative transfer and climate change, as well as measurements of physical constants, and molecular structure. The method is immune to intensity noise of the laser, provides exceptionally long, calibration-free optical pathlengths in a compact volume, and is accurately realized through observations of time and frequency. However, all CRDS studies have involved spectra acquired by the interrogation of one cavity mode at a time [2] or by measuring an unresolved set of cavity modes [3, 4], with most realizations providing no information on dispersion effects. Below, we present a new approach to CRDS incorporating dual phase-locked optical frequency combs for the parallel readout of individual ring-down cavity modes. This is the first approach that exhibits the inherent advantages of single-frequency CRDS while achieving parallel spectral measurements of absorption and/or dispersion.

There have been several demonstrations of broadband laser spectroscopy using steady-state light transmission through cavities. Although these methods do not exhibit the intrinsic noise-immunity of CRDS, they provide simultaneous detection of many species. Applications include trace gas detection in complex gas matrices like human breath [5] and the Earth's atmosphere [6, 7], and observations of complex chemical kinetics [8, 9, 10]. Various schemes to read out cavity transmission probed by an optical frequency comb have been demonstrated, including use of an optical-Vernier coupling approach [11], a swept cavity [12], and cross-dispersive methods [5]. Alternatively, Fourier transform cavity-enhanced spectroscopy has been demonstrated using either dual-comb interferometry [13, 14, 15, 16] or with a mechanically scanned spectrometer [17, 18, 19]. Importantly, these techniques are susceptible to cavity dispersion which causes a mismatch between the probe comb and the comb-like grid of cavity resonances [20, 21].

In CRDS, the modulation of the probe light induces transient fields that oscillate only at the cavity resonant frequencies [22, 23]. Consequently, CRDS decay rates are immune to the probe - cavity frequency mismatch, and signal frequencies caused by parallel cavity excitation encode both mirror and sample dispersion. To measure the lifetimes and frequencies of these component transient fields, the read-out scheme must resolve the individual cavity modes. Although broadband intensity-based, cavity ring-down spectroscopy with an optical frequency comb was demonstrated by Thorpe et al. [4] using an opto-mechanical setup, the spectral resolution was limited to 25 GHz (60 cavity modes). Despite being an elegant demonstration at the time, this proof-of-principle experiment has not evolved into a technology that leverages all the attributes of broadband CRDS.

Here we simultaneously measure spectra comprising multiple resolved cavity modes with approximately 20 kHz linewidth using an optical frequency comb probe and multi-heterodyne detection with a local oscillator (LO) comb, resulting in the first demonstration of dual-comb cavity ring-down spectroscopy (DC-CRDS). Through Fourier analysis of the observed interferograms we measure the widths and positions of cavity resonances, from which absorption and dispersion spectra are derived.

The DC-CRDS technique has the advantages of no instrument line shape, no moving parts, cavity-mode resolution, and fast spectral acquisition of both absorption and dispersion without crosstalk. Along with the experimental results, we present a unified model for the broadband integration of transient cavity response by dual-comb spectroscopy and apply it to the rapid detection of the molecule methane.

## Results

**The idea of DC-CRDS**

Consider an optical cavity excited by an optical frequency comb (probe comb) switched on instantaneously at time $t = 0$ and which beats with another frequency comb (local oscillator) bypassing the cavity. The total output electric field $\tilde{E}_{\text{out}}$ transmitted by the enhancement cavity as a function of time $t \geq 0$ can be described as a sum of fields of individual cavity modes with corresponding teeth of the probe comb [22] and the local oscillator (LO) comb

$$\tilde{E}_{\text{out}} = \sum_j E_{\text{p}_j} \left( e^{i\omega_{\text{p}_j}t} - e^{-\Gamma_{\text{q}_j}t} e^{i\omega_{\text{q}_j}t} \right) + E_{\text{lo}_j} e^{i\left(\omega_{\text{lo}_j}t + \phi_{\text{lo}_j}\right)}, \quad (1)$$

where $E_{\text{p}_j}$ and $E_{\text{lo}_j}$ are field amplitudes of the probe comb at the output of the cavity and LO comb teeth; $\omega_{\text{p}_j}$ and $\omega_{\text{lo}_j}$ are angular frequencies of the probe and LO comb teeth; $\omega_{\text{q}_j}$ and $\Gamma_{\text{q}_j}$ are angular frequencies and spectral widths (HWHM) of the cavity modes, and $\phi_{lo_j}$ are phase shifts between the probe and LO comb teeth. The term $-E_{p_j}e^{-\Gamma_{q_j}t}e^{i\omega_{q_j}t}$ represents the transient response of the cavity (Green's function) induced by the step-change in the probe field amplitude. The beat between the probe and transient response fields describes the cavity buildup (ring-up) signals with oscillations corresponding to their frequency difference $\omega_{p_j} - \omega_{q_j}$. At times $t \gg \Gamma_{q_j}^{-1}$, this transient response term, which contains the cavity resonant frequencies, vanishes and the cavity transmission can be considered to have reached steady state.

The corresponding field $\tilde{E}_{out}$ for the case of instantaneous switching off the probe comb at $t = 0$, is

$$\tilde{E}_{out} = \sum_j E_{p_j}e^{-\Gamma_{q_j}t}e^{i\omega_{q_j}t} + E_{lo_j}e^{i(\omega_{lo_j}t+\phi_{lo_j})}, \qquad (2)$$

where the first term describes fields of conventional ring-down signals of consecutive modes excited by the probe comb and $\tau_{q_j} = \left(2\Gamma_{q_j}\right)^{-1}$ is the conventional intensity-based ring-down time constant. It is clear that the same transient cavity responses, $E_{p_j}e^{-\Gamma_{q_j}t}e^{i\omega_{q_j}t}$, can be observed in both situations — when the probe signal is switched on and off — if the cavity response can be spectrally separated from the probe comb excitation field [23] and from the responses of other cavity modes. For this purpose, a heterodyne beat signal between the comb-like transmission spectrum of the cavity and the LO comb can be observed where the difference in repetition frequencies between the probe and LO combs, $\delta f_r$, must be large enough to resolve the mode widths $\Gamma_{q_j}$.

The down-converted frequency- and time-dependent intensity signal, $I$, of DC-CRDS is schematically presented in Fig. 1 for the case of rapidly switched (on/off) probe comb intensity. The cavity responds with ring-down decays to this alternating switching with a modulation period $T_m$. The Fourier spectra of the heterodyne ring-down-LO beat signals can be easily spectrally separated from probe-LO beat signals which are relatively narrow and have exactly known frequencies. The Fourier transform power spectrum of this time-dependent signal is shown at the top right of Fig. 1a for the coherently and incoherently driven cases of cavity excitation. In the limit of regularly occurring cavity response fields, obtained at the modulation rate, $f_m = T_m^{-1}$, one obtains a comb-like spectrum with teeth spacing of $f_m$ with an envelope corresponding to the cavity mode shape. For random occurrences of cavity excitation, caused by amplitude or frequency noise, the cavity response spectrum is continuous and has a smaller amplitude that is consistent with averaging over an incoherently driven process. The adjacent spectrum of the probe-comb beat signal remains spectrally narrow in both cases, because the comb-cavity phase noise does not influence the degree of mutual coherence between the probe and LO combs. As discussed in **Methods**, the absorption and dispersion spectrum of the intracavity sample are obtained from the halfwidths of the cavity modes and their positions relative to the known comb teeth frequencies, which are manifest in the Fourier spectra.

For comparison between our dual-comb and CW-laser single-mode approaches, on the left side of Fig. 1a we show the unresolved beating intensity (blue) of the probe light and the cavity response for one cavity mode. It reveals the characteristic cavity buildup oscillations exploited before in dispersion spectroscopies [24, 23], followed by ring-down decay. These intra-mode buildup oscillations produce numerous low-frequencies which are observed near zero radiofrequency in the down-converted dual-comb Fourier spectrum, and therefore are well separated from the useful part of the FT spectrum in DC-CRDS.

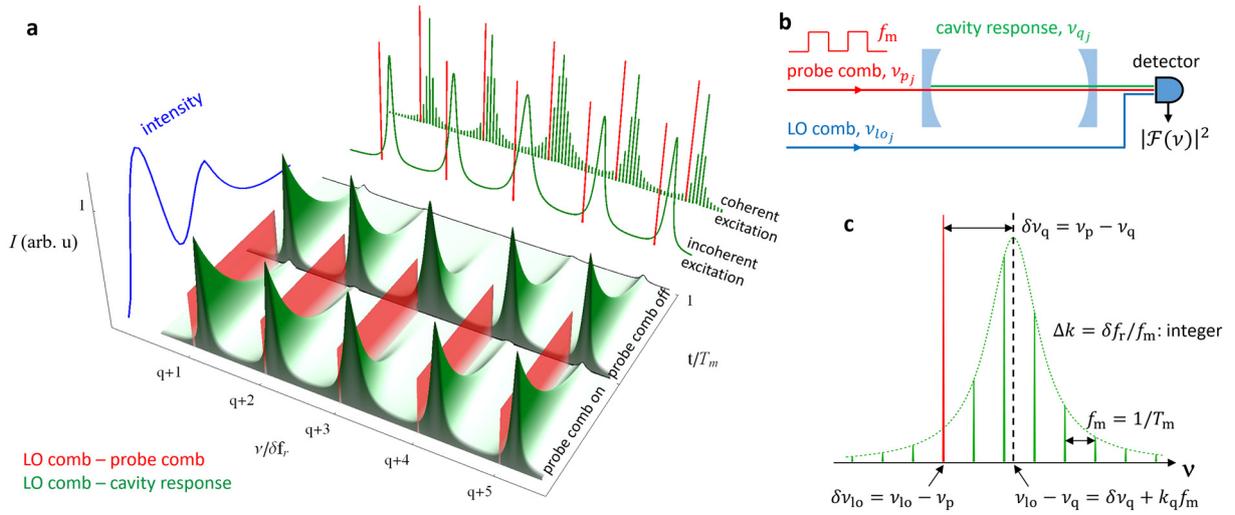

**Fig. 1. Idea of dual-comb cavity ring-down spectroscopy. a,** A simulation of the frequency- and time-dependent intensity of the cavity output/LO comb beat signal is plotted in the spectral range corresponding to the down-converted radiofrequency signal. The cavity responds with ring-down decays (green) to the alternating switching of the probe comb (red) intensity with a square-wave modulation of period $T_m$. For clarity of presentation, the relative intensities between the probe and decay signals are not preserved, and only one period of modulation is shown. Fourier-transform power spectra $|\mathcal{F}(\nu)|^2$ of the probe comb and cavity response are shown at the top right for the coherently driven and incoherently driven cavity excitation cases, corresponding to the acquisition of decay signals at fixed (phase-locked) or random (non-phase-locked) time intervals, respectively. The spectrally unresolved oscillating buildup and ring-down intensity signal (blue), resulting from beating between a single mode and its exciting comb tooth is shown on the left. **b,** A simplified scheme of the experimental setup to generate a signal shown in panel a. **c,** An expanded view of the spectrum of one coherently driven cavity mode illustrating the dominant signal frequencies occurring at multiples of $f_m$ with an intensity envelope corresponding to the mode shape. The diagram corresponds to the present case where $\delta f_r / f_m = \Delta k$ is integer-valued and the total acquisition time is an integral number of modulation periods, $T_m$. With these conditions met, $\nu_{lo} - \nu_p = k_q f_m$, where $k_q$ and $k_{q+1}$ are integer pairs that satisfy $\Delta k = k_{q+1} - k_q$, so that there is no accumulated phase difference between the LO – probe beat signal and the buildup/ring-down cycles. This condition also ensures that the signal at each indicated frequency resides at the respective zeros of all other transform-limited line shapes, thus preventing distortion of the sampled mode shape. This result can be compared to the well-known formula for the frequencies of an optical frequency comb, for the case where the carrier-envelope offset frequency is 0 and where $f_m$, and $k_q$ are analogous to the pulse repetition rate and comb-tooth order respectively.

**Experimental setup**

Our dual-comb cavity ring-down experimental setup is schematically shown in Fig. 2a. Two optical frequency combs (OFC) are generated from a continuous-wave (CW) laser by a set of electro-optic modulators (EOM), similar to those described in Ref. [25]. This single-laser, dual-comb system uses the approach demonstrated in [14] and similarly exhibits high mutual coherence. The output powers of the CW laser, probe- and LO-comb are 8 mW, 30 μW and 30 μW, respectively. Both, probe and LO combs can be independently switched on/off and frequency-shifted by acousto-optic modulators (AOMs). The CW laser frequency is locked to one of the ring-down cavity modes using the Pound-Drever-Hall (PDH) scheme. Each comb has an optical bandwidth of 22 GHz and is centered one free spectral range ($\nu_{FSR}$ = 250 MHz) of the ring-down cavity away from the locked CW laser frequency. The repetition rate (teeth spacing) of the probe comb, $f_r$ = 1 GHz, is matched to four times $\nu_{FSR}$ and the offset frequency, $f_o$, is adjusted by the AOM to roughly match the comb teeth to every fourth cavity mode. Exact comb-cavity frequency matching is not necessary for ring-down measurements nor is it possible when narrow-band molecular dispersion lines are expected in the intracavity sample. In our case frequency matching within $\pm 4$ mode halfwidths was used, but in general the available detuning will depend on the probe comb power. The LO comb has a repetition rate $f_r + \delta f_r$, where $\delta f_r$ = 200 kHz is optimized for the efficient separation of cavity modes, having linewidths of 14 kHz to 20 kHz (FWHM) in the down-converted Fourier spectrum. The LO comb field is combined with the cavity output field to produce a heterodyne beat signal, which is measured by a photodetector (PD) of bandwidth 10 MHz and digitized by an analog-to-digital converter (ADC) with 14 bits of vertical resolution. All RF signals were synchronized to a common reference.

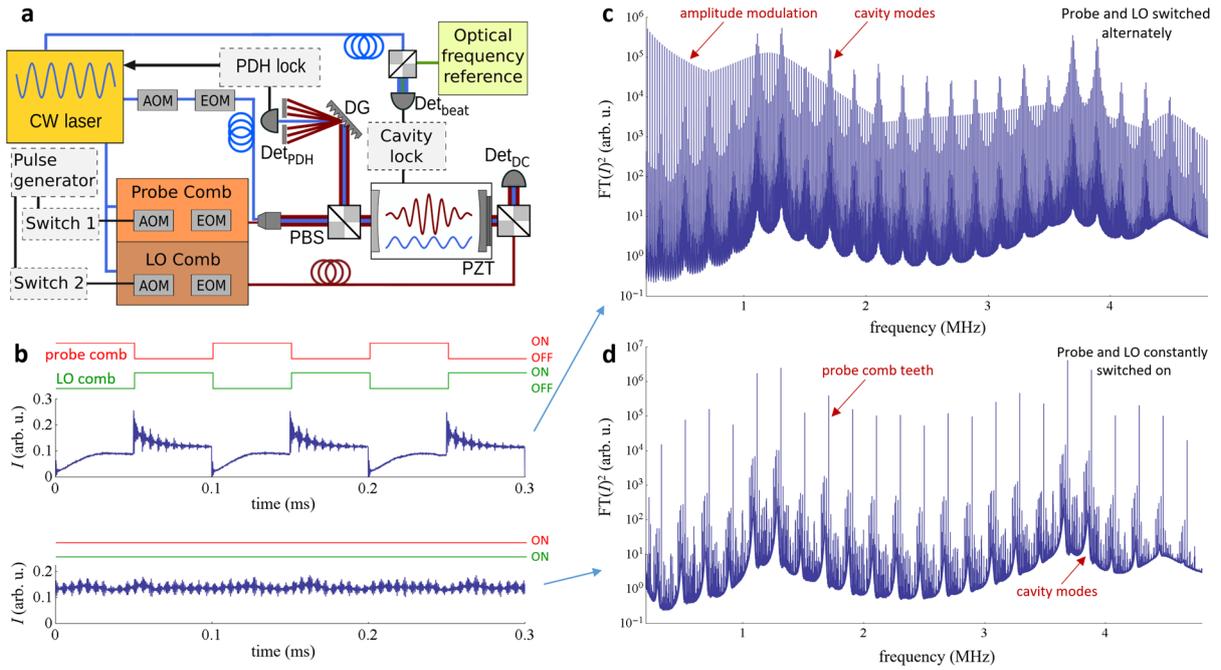

**Fig. 2. Experimental setup of dual-comb cavity ring-down spectrometer. a,** Scheme of experimental setup. Two frequency combs, probe and LO, are generated by a set of electro-optic modulators (EOM) from CW laser locked to the ring-down cavity with the Pound-Drever-Hall (PDH) scheme. The diffraction grating (DG) separates the CW-laser from the comb light. Frequencies of both combs can be shifted by the acousto-optic modulators (AOM) to match the probe comb teeth to the cavity modes and independently shift the LO comb. The ring-down cavity length is stabilized to the femtosecond frequency comb, used as an optical frequency reference. **b,** Light intensity on the detector for the cases of: switching both comb intensities (switches 1 and 2) with square waves having the same frequency $f_m$ = 10 kHz and opposite phases (red and green waveforms), or for both the probe and the LO comb constantly switched on. The beat note of the LO-comb and the cavity output occurs only during the cavity ring-down phase of the modulation in the former case and continuously in the latter case. **c,** The magnitude squared of the Fourier spectrum of the output signal similar the upper one in panel b. The average of 200 spectra, each obtained from a 50-ms-long waveform (500 modulation periods) is shown. Coherently driven ring-down modulation results in a spectrum of cavity modes sampled discretely at multiples of the modulation frequency, $f_m$. Additionally, a square-wave modulation spectrum is visible at odd multiples of $f_m$. **d,** The magnitude squared of the Fourier spectrum of the output signal similar the lower one in panel b. Incoherently driven ring-down and buildup signals result in a continuous spectrum of the cavity modes. The narrow probe comb teeth are also visible.

An example heterodyne beat signal waveforms are shown in Fig. 2b with the ring-down cavity filled with 19.6 kPa (147 Torr) of methane and mixed with 79.1 kPa (593 Torr) of nitrogen. Switching on and off the probe comb field with a square wave at a frequency $f_m$ leads to periodic intensity buildup and ring-down events at the cavity output. The LO comb can be switched on for the desired part of the

waveform – ring-down, buildup or both, to record the resulting beat signal when combined with the cavity output. The Fourier spectrum measured with the LO comb switched on during the ring-down phase is shown in Fig. 2c. A single waveform of duration equal to 50 ms corresponds to a spectral resolution of 20 Hz. With a modulation period of $T_\mathrm{m} = f_m^{-1}$ = 100 µs for the probe comb and phase stabilization between $f_\mathrm{m}$ and $\delta f_r$, the regularity of the time intervals between the ring-down decays leads to coherently averaged cavity mode spectra that are discretely sampled by a comb of frequencies with 10 kHz spacing. The additional resonances which are visible at odd multiples of 10 kHz are caused by the square-wave amplitude modulation of the probe comb.

Interestingly, when both the probe and LO comb fields are constantly switched on, the Fourier spectrum, shown in Fig. 2d also reveals cavity mode shapes. They originate from residual phase/frequency and amplitude perturbations in the probe laser. These transient events occur randomly in time and induce decaying mode fields of random phase. As a result, these aperiodic signals add incoherently and yield a continuous spectrum of cavity modes that are clearly visible at the bottom of the spectrum. We note the presence of a similar continuous mode spectrum contained in the lower envelope of Fig. 2c, which is more than four orders of magnitude weaker than the dominant coherent comb-like spectrum, and which we attribute to residual frequency or amplitude noise in the locked laser frequency. As discussed in **Methods**, the amplitudes of these continuous spectra were observed to depend on the magnitude of the frequency noise between the probe comb and the cavity, consistent with our explanation for the signal origin. The additional strong resonances correspond to the steady-state beating between the probe and local oscillator combs which are used in conventional dual-comb spectroscopy [26, 27, 13, 14], but which are not used in the present analysis.

**Molecular spectra retrieval**

In the case of coherently driven ring-down signals, individual Fourier spectra were calculated from time periods $T_\mathrm{m}$ = 100 µs, corresponding to one cavity excitation and ring-down event and averaged. The straightforward selection of frequencies corresponding to even multiples of $f_\mathrm{m}$ from the Fourier spectrum leads to a clean spectrum of the cavity modes, shown as black dots in Fig. 3a. Although the time period, $T_\mathrm{m}$, leads to 10 kHz spacing between spectral points, the transform-limited line shape of the Fourier spectrum (sinc function) was cancelled because $T_\mathrm{m}$ was set to an integer multiple of $(\delta f_\mathrm{r})^{-1}$ [28, 19, 29]. Each cavity mode was fitted with an asymmetric Lorentzian shape with a linear background, given in Eq. (7) in **Methods**. The average signal-to-noise ratio of the mode shapes was 330, calculated as the ratio of the mode amplitude to the standard deviation of the fit residuals shown at the bottom of Fig. 3a.

From the fitted mode halfwidths and positions we calculated the methane absorption and dispersion spectra shown as black dots in Fig. 3b. The red line corresponds to the simulated methane spectrum with line positions, relative intensities and pressure broadening parameters taken from the reference HITRAN2016 database [30]. The overall intensity of both the absorption and dispersion parts of the simulated spectrum and their individual linear backgrounds were fitted to the experimental data. For the absorption spectra, the background represents the cavity mirror losses, while for dispersion case the linear background results from mismatch between the averaged cavity free spectral range $\nu_\mathrm{FSR}$ and the probe comb repetition rate $f_\mathrm{r}$, and from the constant detuning between the probe comb and cavity modes set by the comb offset $f_\mathrm{o}$. The shapes of the measured and calculated spectra are in good agreement.

The ratio of the fitted methane spectrum intensity to the HITRAN2016 reference value is 0.96 with a standard deviation of 0.05. Given the uncertainties in the reference data (< 10 % for line intensities and > 20 % for line widths) [30] and in methane mole fraction (5 % associated with independently measured time-dependent mixing of methane with nitrogen in the ring-down cavity), the agreement between database and experiment is very good, providing evidence that the measurements are not subject to significant bias. The absorption and dispersion spectra are also in good mutual agreement as shown in Fig. 3b.

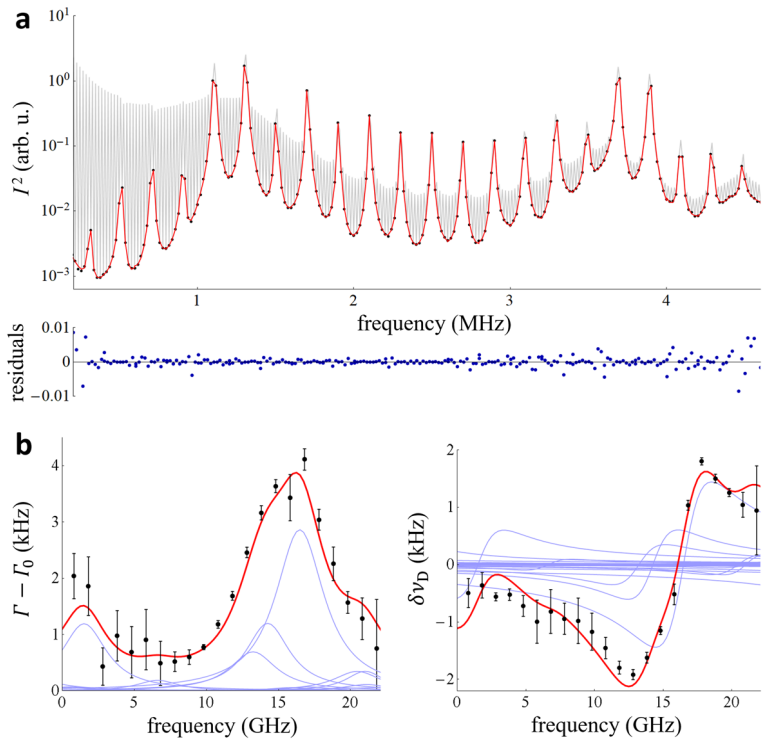

**Fig. 3. Molecular spectra measured by coherently driven cavity excitation. a,** Fourier spectrum of the down-converted ring-down signals when the LO comb was switched on for the ring-down phase only (gray line). An average of 1500 spectra based on consecutive 100-$\mu$s-long ring-down signals is shown. Selection of even multiples of $f_m$ leads to clean cavity mode spectra comprising 10 points per mode (black points). The fitted mode shapes are shown as red lines. Below, the ratio of the fit residuals to the mode peaks is shown as blue points. **b,** Absorption (left plot) and dispersion (right plot) spectra of methane (black points) retrieved from the measured cavity mode widths and positions. An offset of 191274.65 GHz has been subtracted from the frequency axis, and the error bars correspond to three times the measured ensemble standard deviation at each spectrum detuning. The methane spectra calculated using line shape parameters from HITRAN2016 are shown as the red lines with individual transitions shown as blue lines. The overall intensity was fitted simultaneously to the measured absorption and dispersion spectra. For the absorption spectrum, the fitted background, corresponding to the empty cavity losses of 6.30(12) kHz, was subtracted from the plot. Likewise, for the dispersion case, the fitted linear background was subtracted from the plot.

Analysis of the relatively weak modes observed in the continuous spectra requires their separation from the much stronger and narrow resonances found in the full Fourier spectrum shown in Fig. 2d. To this end, we averaged Fourier spectra of waveforms 50 ms in duration, resulting in a resolution of 20 Hz. Selected points of the spectrum are shown in Fig. 4a together with fitted asymmetric Lorentzian mode shapes with a linear background (Eq. (7) in Methods). In the inset of Fig. 4a a zoom on one cavity mode is shown. The strong resonances were removed from the continuous spectrum as follows: first, a rough removal of every point higher by 50 % than the initially expected mode intensity and the initial fit; second, removal of every point having initial fit residuals higher than the absolute value of the lowest (negative) residual point and the final fit. This two-step fitting led to the flat fit residuals shown in the bottom of Fig. 4a and its inset. The average signal-to-noise ratio of the mode shapes was 45:1, nearly an order of magnitude lower than that obtained for the coherent excitation case. However, because the continuous incoherently driven spectrum has 1000 times more points per mode width than the coherently driven DC-CRDS, the standard deviations of the fitted mode halfwidths and positions are lower for the continuous spectra, as expected.

The absorption and dispersion methane spectra, obtained from the mode widths and shifts of continuous mode spectra are shown in Fig. 4b. We compare the measured spectra with the reference HITRAN2016 data in the same fashion as for the coherent excitation results presented earlier. In this case, the ratio of fitted to reference spectra intensity and its standard deviation is 0.92 ± 0.02, which is within the stated uncertainty of the reference data and our sample mixing ratio, as discussed earlier. The spectrum intensities fitted from the coherent and incoherent excitation approaches agree to within combined standard uncertainties.

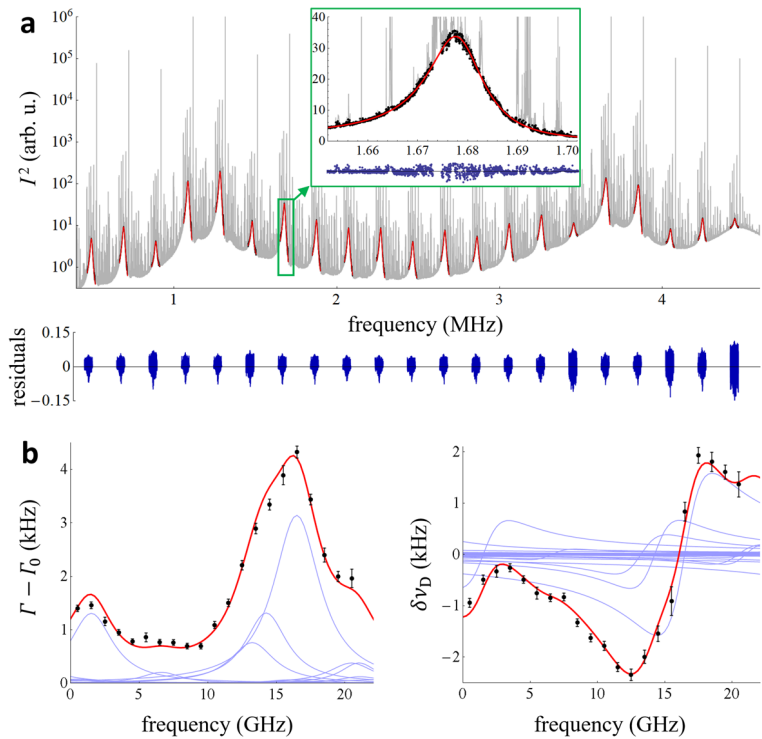

**Fig. 4. Molecular spectra measured by incoherently driven cavity excitation. a,** Fourier spectrum of the down-converted probe comb continuously transmitted through the cavity and ring-down events excited by phase noise of the probe comb (gray line). An average of 2000 spectra of consecutive signals each 50 ms in duration is shown. In the inset, a zoom of one cavity mode is shown. Black points are the data selected for mode shape fitting and the fitted mode shapes are shown as red lines. Below, ratios of the fit residuals to the mode peaks are shown as blue lines. **b,** Absorption (left plot) and dispersion (right plot) spectra of methane (black points) retrieved from the measured cavity mode widths (HWHM) and positions. An offset of 191273.96 GHz has been subtracted from the frequency axis, and the error bars correspond to the three times the measured ensemble standard deviation at each spectrum detuning. The reference shape of the methane spectrum from HITRAN2016 is shown as a red line with individual transitions shown as blue lines. The reference spectrum overall intensity was fitted simultaneously to the absorption and dispersion experimental spectra. The fitted absorption spectrum background of 6.10(5) kHz, corresponding to the empty cavity losses was subtracted from the plot. The fitted dispersion spectrum linear background representing a mismatch between the average cavity FSR and the probe comb $f_r$ was also subtracted from the plot.

## Discussion

Our novel dual-comb cavity ring-down spectroscopy (DC-CRDS) combines the advantages of continuous-wave CRDS with parallel measurements of absorption and dispersion on many cavity modes. Similar to conventional CRDS, the ultimate spectral resolution is limited by the widths of the cavity modes, and the cavity length determines the molecular spectrum point spacing in a single parallel measurement. The measured spectra are highly insensitive to the temporal and spectral variations of the dual-comb intensity. This property is advantageous compared to conventional intensity-based dual-comb spectroscopy, which requires calibration for the comb power spectrum and sample-dependence of the comb transmission caused by dispersive shifts of the cavity resonances. It is worth noting that in the spectrum calculation there was no need to account for the 100-fold variation in the power of our comb teeth over the 22-GHz spectral range considered. This wide range of power in the probe comb did not lead to observable systematic biases. The immunity to temporal power variation is also an advantage compared to recently developed comb-based broadband cavity mode width spectroscopy [31, 32, 33] using step-scanning of the mode shapes.

The DC-CRDS requirements for the comb-cavity phase/frequency noise and for matching the resonant frequencies with the probe comb frequencies are relaxed compared to other comb-based cavity-enhanced spectroscopies. Although the phase/frequency noise does decrease the ratio of coherently driven to incoherently driven cavity excitation efficiency, we have demonstrated that both approaches can be used for reliable measurements of the molecular spectra. The case of coherently driven excitation enables relatively fast spectrum acquisition with moderate light intensity. In contrast, the incoherently driven approach is less technically demanding in terms of locking the probe to the cavity, but requires longer averaging times to achieve comparable precision. In both cases the width of the relative LO comb - cavity frequency noise must be small compared to the cavity mode width to avoid instrumental spectrum broadening.

Apart from the bandwidth of the dual-comb itself and reductions of signal intensity caused by cavity mirror dispersion, the bandwidth of the simultaneously measured optical spectrum $\Delta\nu_{\text{opt}}$ for DC-CRDS is limited by the maximum cavity mode width $\Gamma$, comb mode spacing $f_r$ and optical detector bandwidth $\Delta f_d$. Assuming that mode separations of $m \times \Gamma$ are sufficient to resolve and precisely fit mode widths and shifts, the minimum $\delta f_r = m\Gamma$, whereas the maximum optical bandwidth is

$$\Delta\nu_{\text{opt}} = \begin{cases} \frac{\Delta f_d}{m\Gamma} f_r & \text{for} \quad f_r > 2\Delta f_d \\ \frac{1}{2m\Gamma} f_r^2 & \text{for} \quad f_r < 2\Delta f_d \end{cases},$$

where the case of $f_r > 2\Delta f_d$ corresponds to the detector-bandwidth-limited condition, with the second case being the comb-repetition-rate limit. Because of the need to measure $\Gamma$, which remains unchanged in the optical-to-RF down-conversion process, the achievable optical bandwidth is lower than that of conventional dual-comb spectroscopy. Nevertheless, appreciable optical bandwidths commensurate with characteristic widths of molecular absorption bands are possible with DC-CRDS. For example, with a readily available fast detector and digitizer, $\Delta f_d = 200$ MHz, and high finesse cavity, $\Gamma = 1$ kHz, for $f_r = 1$ GHz, and mode separation $m = 10$ half-widths, one obtains a bandwidth limit $\Delta\nu_{\text{opt}} = 20$ THz. A higher bandwidth DC-CRDS spectrum could be measured sequentially by adjusting the probe comb frequencies to compensate for cavity mirror dispersion.

While we presented a spectral bandwidth of 22 GHz, limited by the dual-comb system in our proof-of-principle experiment, more broadband femtosecond and electro-optic comb systems are available in many wavelength ranges, from the visible to mid-IR regions (see e.g. [34, 35, 36, 37, 38, 39]). Because of the relaxed requirements for the relative phase stability of the probe comb and cavity discussed above, we also expect that the DC-CRDS measurements should be possible using pairs of femtosecond-based optical frequency combs, even without long-term phase locking between the two combs. To prevent instrumental broadening of the cavity modes in both the coherently and incoherently driven cases, only the LO comb must be locked to the cavity, whereas the quality of the probe comb lock to the cavity affects only the relative intensities of the coherently and incoherently driven contributions to the Fourier spectrum. However, the incoherently driven approach requires averaging to cancel the phase-dependent instrumental lineshape (sinc) function.

The combination of spectral measurement times in the ms range, sensitivity, accuracy, specificity, spectral coverage, mode resolution and experimental simplicity demonstrated here by DC-CRDS is potentially transformative. This is a new approach to enable accurate, massively parallel and rapid measurements of weakly absorbing species. Here, as a test case, we study the spectrum of methane, having an absorption coefficient below $2 \times 10^{-6}$ cm$^{-1}$, in an important water-window for the remote sensing of the Earth's atmosphere. We anticipate that follow-up DC-CRDS measurements throughout the near- and mid-infrared will yield improvements in methane spectral parameters that are critical to identifying natural gas leaks [40], enabling precision radiative transfer and climate change models [41] and spectroscopy of foreign-broadened methane relevant to the search for exoplanet companion biosignatures [42, 43]. Other applications include non-invasive optical sensors to probe complex gas matrices such as breath [44], as well as agile multispecies trace gases analyzers for atmospheric composition monitoring [45, 46], characterisation of broadband mirror loss and dispersion [47, 48], chemical reaction kinetics [8] and collisional processes in gases [49]. Finally, DC-CRDS could bring all advantages of CRDS to frontier field of research on high-resolution broadband spectroscopy of cold large organic molecules [50] and other complex molecular systems [51].

## Methods

### Model of the Fourier spectrum

Similar to the derivation of the cavity buildup spectrum given in Ref. [23], the intensity of the DC-CRDS field given by Eq. (1), is $I(t) = \left|\text{Re}\{\tilde{E}_{\text{out}}(t)\}\right|^2$. Assuming ideal frequency locking of the probe comb to the cavity and no amplitude modulation other than switching on and extinction of the probe beam, for the simple case of one cavity mode excited by the probe field at $t = 0$ and extinguished at time $t_s$, the buildup signal in the interval $0 \leq t \leq t_s$ is,

$$I_{\text{b}}(t) = I_{\text{p}}\left[1 + e^{-2\Gamma_q t} - 2e^{-\Gamma_q t}\cos(\delta\omega\, t)\right] + I_{\text{lo}} + 2\sqrt{I_{\text{p}}I_{\text{lo}}}\left[\cos(\delta\omega_{\text{lo}}\, t + \phi_{\text{lo}}) - e^{-\Gamma_q t}\cos\left((\delta\omega_q + \delta\omega_{\text{lo}})t + \phi_{\text{lo}}\right)\right], \quad (3)$$

and from Eq. (2) the decay signal for $t > t_s$ is

$$I_{\text{d}}(t) = I_{\text{lo}} + I_{\text{p}}\, e^{-2\Gamma_q(t-t_s)} + 2\sqrt{(I_{\text{lo}}I_{\text{p}})}e^{-\Gamma_q(t-t_s)}\cos\left((\delta\omega_q + \delta\omega_{\text{lo}})(t - t_s) + \phi_{\text{lo}}\right), \quad (4)$$

in which $\delta\omega_q = \omega_p - \omega_q$, $\delta\omega_{\text{lo}} = \omega_{\text{lo}} - \omega_p$, $I_p = E_p^2/2$, $I_{\text{lo}} = E_{\text{lo}}^2/2$ and $\phi_{\text{lo}}$ is the phase shift between the local-oscillator and probe combs. The Fourier transform of $I_b(t)$ gives a sum of seven complex Lorentzian resonances corresponding to

$$\mathcal{F}_{\text{b}}(\omega) = I_p \left\{\frac{1}{2\Gamma_q - i\omega} - \left[\frac{1}{\Gamma_q - i(\omega + \delta\omega_q)} + \frac{1}{\Gamma_q - i(\omega - \delta\omega_q)}\right]\right\} + \sqrt{I_p I_{\text{lo}}}\left\{\left[\frac{e^{i\phi_{lo}}}{\Gamma_{\text{lo}} - i(\omega + \delta\omega_{\text{lo}})} + \frac{e^{-i\phi_{lo}}}{\Gamma_{\text{lo}} - i(\omega - \delta\omega_{\text{lo}})}\right] - \left[\frac{e^{i\phi_{lo}}}{\Gamma_q - i(\omega + \delta\omega_q + \delta\omega_{\text{lo}})} + \frac{e^{-i\phi_{lo}}}{\Gamma_q - i(\omega - \delta\omega_q - \delta\omega_{\text{lo}})}\right]\right\}, \quad (5)$$

where $\Gamma_{lo}$ accounts for the relative probe comb – LO comb halfwidth caused by variation in $\phi_{lo}$. When neglecting this phase variation, $\Gamma_{lo} = 0$ and these resonances reduce to delta functions. The set of resonances at low frequencies, $\omega = 0$ and $\delta\omega_q$, are not resolved in parallel cavity mode excitation. For DC-CRDS the relevant resonances are those at $\omega = \delta\omega_q + \delta\omega_{lo}$, because $\delta\omega_{lo}$ increases by $2\pi\,\delta f_r$ for successive excited cavity modes. Similarly, the Fourier transform of the decay signal, $I_d(t)$ results in the following three complex Lorentzian resonances

$$\mathcal{F}_d(\omega) = I_p \left\{ \frac{1}{2\Gamma_q - i\omega} \right\} + \sqrt{I_p I_{lo}} \left\{ \frac{e^{i\phi_{lo}}}{\Gamma_q - i(\omega + \delta\omega_q + \delta\omega_{lo})} + \frac{e^{-i\phi_{lo}}}{\Gamma_q - i(\omega - \delta\omega_q - \delta\omega_{lo})} \right\}, \quad (6)$$

which does not include the four resonances at $\pm\delta\omega_q$ and $\pm\delta\omega_{lo}$, because of the absence of the probe field.

In practice, Eqs. (5) and (6) can be used to calculate the real-valued quantity, $|\mathcal{F}(\omega)|^2$, to model the power spectral density of the measured signal for each time interval. Evaluation of $|\mathcal{F}(\omega)|^2$ accounting for cross-terms and wings from non-local resonances can be approximated by Lorentzian shapes with asymmetric, dispersive components [23] that are dependent on the phase $\phi_{lo}$.

Using the preceding model for DC-CRDS Fourier spectra, we retrieve the mode widths and positions by fitting mode spectra near their centers with asymmetric Lorentzian shapes and a linear background approximating the joint contributions of distant modes

$$|\mathcal{F}(\omega)|^2 \approx a \frac{\Gamma_q^2 \left(1 + y(\omega - \delta\omega_q - \delta\omega_{lo})\right)}{(\omega - \delta\omega_q - \delta\omega_{lo})^2 + \Gamma_q^2} + b(\omega - \delta\omega_q - \delta\omega_{lo}) + c, \quad (7)$$

where $y$ is an asymmetry parameter, and $a$, $b$ and $c$ are the mode amplitude and two linear background parameters, respectively.

**Generalized cavity response with frequency and amplitude modulation**

Consider a single cavity mode, $q$, of angular frequency $\omega_q$, and field decay rate, $\Gamma_q$, probed by a time-dependent field, $\tilde{E}_p(t)$ from one tooth of an optical frequency comb which exhibits step changes, denoted by, $j$, in amplitude $\Delta E_j = E_j - E_{j-1}$ and/or angular frequency $\omega_{p,j} - \omega_{p,j-1}$. We also define $\bar{\omega}_p = \omega_q + \delta\omega_q$ as the average angular frequency of the probe field and specify $d\omega_{p,j} = \omega_{p,j} - \bar{\omega}_p$ as the frequency deviation from the mean value. During each step beginning at time, $t = t_j$, the amplitude and frequency are constant over the interval $\Delta t_j$. With these specified stepwise changes in the probe field, the net field exiting the ring-down cavity equals the sum of the probe fields over the step intervals,

$$\tilde{E}_p(t) = e^{i\bar{\omega}_p t} \sum_j \text{rect}(z_j) \tilde{\eta}_j E_j e^{id\omega_{p,j} t} = \tilde{f}(t) e^{i\bar{\omega}_p t}, \quad (8)$$

and the corresponding sum of cavity mode fields (induced by the transient responses of the cavity to the time-dependent probe field) is

$$\tilde{E}_q(t) = -\sum_j \Delta\tilde{E}_j e^{(i\omega_q - \Gamma_q)(t - t_j)} H(t - t_j) = -\tilde{g}(t) e^{(i\omega_q - \Gamma_q)t}. \quad (9)$$

Here, rect($z_j$) is the unit rectangle function which is nonzero only on the interval -1/2 to 1/2, $z_j = \frac{(t - t_j)}{\Delta t_j} - \frac{1}{2}$, $\tilde{\eta}_j = [\Gamma_q/([\Gamma_q - i(\bar{\omega}_p + d\omega_{p,j} - \omega_q)]$ which accounts for the detuning dependence of the steady-steady coupling efficiency of the probe laser field into the cavity, and $H(t)$ is the Heaviside step function. The complex-valued change in the field at each step $j$ inducing the mode fields is given by

$$\Delta \tilde{E}_j = e^{i\bar{\omega}_p\, t_j}\{\tilde{\eta}_j E_j e^{id\omega_{p,j}\, t_j} - \tilde{\eta}_{j-1} E_{j-1} e^{id\omega_{p,j-1}\, t_j}\}. \quad (10)$$

The local oscillator field of the nearest comb tooth, $\tilde{E}_{lo}(t)$, has an angular frequency shifted relative to the probe field by $\delta\omega_{lo}$, and can be written as

$$\tilde{E}_{lo}(t) = \tilde{h}(t) e^{i(\bar{\omega}_p + \delta\omega_{lo})t + \phi_{lo}}, \quad (11)$$

where $\tilde{h}(t)$ is proportional to $\sum_j \text{rect}(z_j) E_j e^{id\omega_{p,j}\, t}$ and we have assumed that $\phi_{lo}$ is a constant phase difference between the local oscillator and probe fields.

The sum of the probe, mode and local oscillator fields incident on the photodetector becomes,

$$\tilde{E}_p(t) + \tilde{E}_q(t) + \tilde{E}_{lo}(t) = \tilde{f}(t) e^{i\bar{\omega}_p t} - \tilde{g}(t) e^{(i\omega_q - \Gamma_q)t} + \tilde{h}(t) e^{i(\bar{\omega}_p + \delta\omega_{lo})t + \phi_{lo}}, \quad (12)$$

in which the functions $\tilde{f}(t), \tilde{g}(t),$ and $\tilde{h}(t)$ account for time-dependent variations in the incident probe field amplitude and frequency. The resulting fields and Fourier spectra of the heterodyne beat signals can be readily determined from Eqs. (8)-(12) for specified deterministic or stochastically varying amplitudes and/or frequency variations in the probe field. Importantly, the net cavity time response generally contains a term proportional to $e^{(i\omega_q - \Gamma_q)t}$, which when combined with the local-oscillator field results in resonances at $\omega_q = \pm (\delta\omega_q + \delta\omega_{lo})$. This general result captures how the cavity modes can be excited by coherently driven effects or by random perturbations in the amplitude and/or frequency of the probe field.

**Sensitivity and speed of DC-CRDS**

The absorption coefficient of the sample at the $q$-th cavity mode, $\alpha_q$, centered at frequency $\nu_q$, is proportional to the change of the cavity mode halfwidth (HWHM), $\Delta\Gamma_q = \frac{c}{4\pi}\alpha_q$, compared to the empty-cavity mode width $\Gamma_0$. Here $c$ is the speed of light. $\Delta\Gamma_q$ is linked to the dispersive cavity mode shift $\delta\nu_D$ by the complex-valued refractive index $n(\nu) = n_0 + \frac{\chi(\nu)}{2n_0}$, where $n_0$ is the non-resonant refractive index and the resonant susceptibility $\chi(\nu) = \chi'(\nu) - i\chi''(\nu)$ [22]. For an isolated spectral line, the relation between $\Delta\Gamma_q$ and $\delta\nu_D$ can be written using the complex-valued line shape function $\mathcal{L}(\nu)$ [52, 53]

$$\frac{\delta\nu_D}{\Delta\Gamma_q} = -\frac{\chi'}{n_0 \chi''} = \frac{\text{Im}[\mathcal{L}(\nu_q - \nu_0)]}{n_0 \text{Re}[\mathcal{L}(\nu_q - \nu_0)]}. \quad (13)$$

To demonstrate the speed and sensitivity of our coherently driven DC-CRDS realization, in Fig. 5 we present Allan deviations of fitted mode halfwidths and positions averaged over all 22 simultaneously measured modes shown in Fig. 3. Both, the halfwidth and position have similar sensitivities of 1.3 kHz in 2 ms averaging or 70 Hz to 100 Hz in 1 s for width and positions, respectively. For each spectral element (cavity mode), these values correspond to noise-equivalent absorption coefficients of $2.94 \times 10^{-8}$ cm$^{-1}$Hz$^{-1/2}$ and $4.2 \times 10^{-8}$ cm$^{-1}$Hz$^{-1/2}$ from mode widths and positions, respectively. When both widths $\Delta\Gamma_q$ and positions $\delta\nu_D$ are used, as done in Fig. 3b one obtains $2.6 \times 10^{-8}$ cm$^{-1}$Hz$^{-1/2}$ and $5.5 \times 10^{-9}$ cm$^{-1}$Hz$^{-1/2}$ per spectral element. No drift in measured halfwidths or positions was observed for averaging times up to at least 30 s. Small deviations from linearity in the log-log plot for mode position near $t$ = 10 ms may be a result of our cavity length stabilization bandwidth which is below 100 Hz.

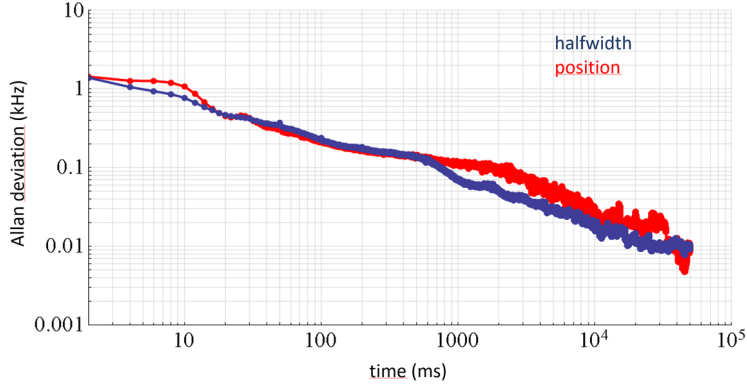

**Fig. 5. Allan deviations of fitted half widths and positions of cavity modes.** The plotted Allan deviations of fitted halfwidth (blue) and position (red) is averaged over all 22 simultaneously measured modes in the coherently driven excitation approach. One set of cavity modes was obtained from an average of 20 Fourier spectra based on individual signals that were 100 $\mu$s in duration.

**Details of molecular spectra fitting**

The absorption $\Gamma(\nu)$ and dispersion $\delta\nu(\nu)$ parts of the methane spectra, shown in Figs. 3b and 4b, were fitted simultaneously with a model given by

$$\Gamma(\nu) = a \times \Gamma_{\mathrm{HT}}(\nu - \nu_{\mathrm{l}}) + \Gamma_0 , \qquad (14)$$

$$\delta\nu(\nu) = a \times \delta\nu_{\mathrm{HT}}(\nu - \nu_{\mathrm{l}}) + \delta\nu_0 + d\nu_{\mathrm{FSR}} \times (\nu - \nu_{\mathrm{l}}), \qquad (15)$$

where $\Gamma_{\mathrm{HT}}$ and $\delta\nu_{\mathrm{HT}}$ are the absorption and dispersion lineshape models of the spectrum calculated from Voigt profile line shape parameters for the individual methane lines in HITRAN2016 [30]. The fitted parameters are: $a$ – amplitude of the spectrum, $\nu_{\mathrm{l}}$ – position of the spectrum on the local frequency axis, $\Gamma_0$ – mode width corresponding to the broadband cavity losses, $\delta\nu_0$ – constant detuning of the probe comb from the cavity modes, $d\nu_{\mathrm{FSR}}$ – mean difference between $4 \times \nu_{\mathrm{FSR}}$ and the probe comb $f_{\mathrm{r}}$. The amplitude $a$ and position $\nu_{\mathrm{l}}$ were fitted as shared parameters between absorption and dispersion.

**Phase (frequency) noise vs incoherently driven signal amplitude**

The continuous spectrum of the cavity modes, shown in Fig. 2d and Fig. 4a are caused by the random excitation of cavity ring-down signals originating from perturbations in the relative phase/frequency between the probe comb and the intracavity field. Perturbations in the amplitude can also contribute to these random excitation of the cavity modes as discussed above. Because each perturbation in the incident field induces a ring-down response with opposite phase (see Eq. (1)), the magnitude of the continuous cavity mode spectra is expected to depend on the relative phase or frequency noise between the probe comb and the cavity modes. We demonstrate this dependence by comparison of the measured spectra for two magnitudes of the phase noise, which can be regulated by adjusting the gain in the PDH lock of the CW laser to the cavity (see Fig. 2a). The magnitude of this effect is manifest in the PDH lock error signal. In Fig. 6a two continuous spectra of cavity modes are shown. These spectra were measured with the PDH lock gain set to strong and weak phase lock conditions, respectively. This degradation of the PDH lock nearly doubles the phase noise intensity at frequencies below 60 kHz. The corresponding spectra of the PDH error signals are shown in Fig. 6b for both cases. Clearly, the higher comb-cavity phase noise increases the amplitude of the mode spectrum acquired under continuous acquisition conditions.

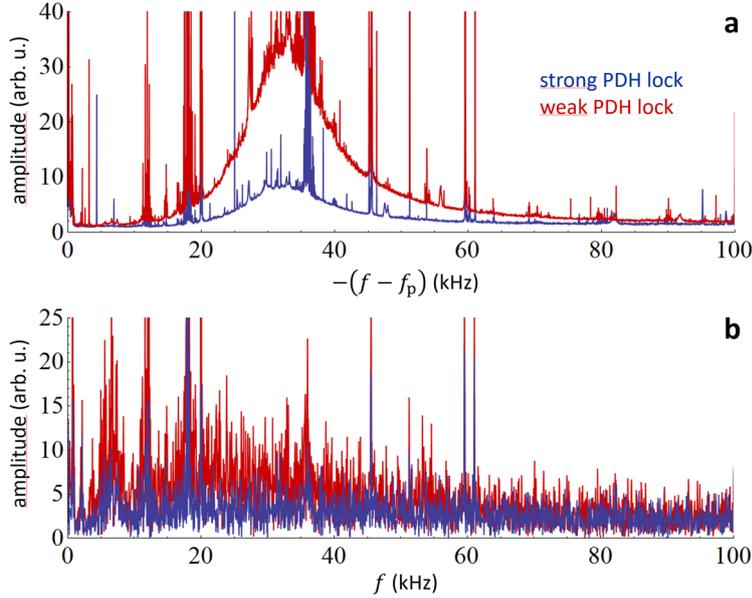

**Fig. 6. Dependence of cavity mode spectrum on the magnitude of comb-cavity phase noise. a,** Spectrum of cavity modes acquired under continuous excitation in the case of strong (blue) and weak (red) PDH lock of the CW laser to the ring-down cavity. The frequency axis is shifted by the frequency of the probe comb tooth $f_p$, which excites the selected mode, and inverted to match the frequency axis of panel b, so that zero corresponds to the probe comb tooth center. **b,** The spectrum of the PDH error signal in the case of a strong (blue) and weak (red) PDH lock, corresponding to the measured modes shown in panel a.

We have also confirmed by numerical simulations that the dual-comb Fourier spectrum of consecutive ring-down signals defined by the field of Eq. (1) with added random phase shift between the comb and cavity response fields leads to continuous cavity mode spectra. Calculations based on Eqs. (8)-(12) also reveal that for non-zero probe-cavity frequency detuning values with phase noise, Fourier spectra that are sampled at non-integral multiples of $f_m$ exhibit modulation because of convolution with the phase-dependent sinc function that corresponds to the transform-limited lineshape. Nevertheless, averaging over these phase-noise-induced fluctuations yields undistorted mode shapes. However, the special case of zero probe comb detuning can result in mode splitting caused by competition between randomly occurring buildup and ring-down events, which have opposite signs.

## Data availability

All data supporting the findings of this study are available from the corresponding author upon reasonable request.

## Acknowledgements

The research was supported by the National Science Centre, Poland project Nos. 2015/18/E/ST2/00585 and 2016/23/B/ST2/00730. AJF and JTH were funded by NIST. The research is part of the program of the National Laboratory FAMO in Toruń, Poland.

## Author contributions

D.L., J.T.H., A.J.F., and A.C. outlined the idea of parallel dynamic mode-resolved heterodyne spectroscopy. D.L. and P.M. designed the concept of dual-comb cavity ring-down spectroscopy experiment. D.C., A.N., G.K., and P.M. built DC-CRDS spectrometer. T.V., T.W., V.B., and T.H. built a setup to generate dual comb from cw laser. D.C. measured experimental data. D.L. analyzed and interpreted experimental data with contribution from J.T.H., R.C., D.C. and P.M. D.L. wrote the original draft of the manuscript and prepared figures. D.L., J.T.H. and R.C. contributed to developing the theory of DC-CRDS. All authors contributed to the final version of manuscript. D.L. and P.M. coordinated the project.

## Competing interests

All authors declare no competing interests.